\documentstyle[onecolumn]{mn}
\def\RCB{$\rho$ CrB}
\def\MJ {$M_{J}$}
\def\ms {m s$^{-1}$}
\def\Mo {$M_{\odot}$}
\def\mas {{\rm milli-arc-sec}}
\def\GHB{Gatewood, Han \& Black (2001)}

\newif\ifAMStwofonts
 
\ifoldfss
  \ifCUPmtlplainloaded \else
    \NewTextAlphabet{textbfit} {cmbxti10} {}
    \NewTextAlphabet{textbfss} {cmssbx10} {}
    \NewMathAlphabet{mathbfit} {cmbxti10} {} 
    \NewMathAlphabet{mathbfss} {cmssbx10} {} 
  \fi
  \ifAMStwofonts
    \ifCUPmtlplainloaded \else
      \NewSymbolFont{upmath} {eurm10}
      \NewSymbolFont{AMSa} {msam10}
      \NewMathSymbol{\upi}     {0}{upmath}{19}
      \NewMathSymbol{\umu}     {0}{upmath}{16}
      \NewMathSymbol{\upartial}{0}{upmath}{40}
      \NewMathSymbol{\leqslant}{3}{AMSa}{36}
      \NewMathSymbol{\geqslant}{3}{AMSa}{3E}

    \fi
  \fi
\fi 
 
\ifnfssone
  \newmathalphabet{\mathit}
  \addtoversion{normal}{\mathit}{cmr}{m}{it}
  \addtoversion{bold}{\mathit}{cmr}{bx}{it}
  \newmathalphabet{\mathbfit} 
  \addtoversion{normal}{\mathbfit}{cmr}{bx}{it}
  \addtoversion{bold}{\mathbfit}{cmr}{bx}{it}
  \newmathalphabet{\mathbfss} 
  \addtoversion{normal}{\mathbfss}{cmss}{bx}{n}
  \addtoversion{bold}{\mathbfss}{cmss}{bx}{n}
  \ifAMStwofonts
    \ifCUPmtlplainloaded \else
      \UseAMStwoboldmath
      \makeatletter
      \new@mathgroup\upmath@group
      \define@mathgroup\mv@normal\upmath@group{eur}{m}{n}
      \define@mathgroup\mv@bold\upmath@group{eur}{b}{n}
      \edef\UPM{\hexnumber\upmath@group}
      \new@mathgroup\amsa@group
      \define@mathgroup\mv@normal\amsa@group{msa}{m}{n}
      \define@mathgroup\mv@bold\amsa@group{msa}{m}{n}
      \edef\AMSa{\hexnumber\amsa@group}
      \makeatother
      \mathchardef\upi="0\UPM19
      \mathchardef\umu="0\UPM16
      \mathchardef\upartial="0\UPM40
      \mathchardef\leqslant="3\AMSa36
      \mathchardef\geqslant="3\AMSa3E
    \fi
  \fi
\fi 
 
\ifnfsstwo
  \DeclareMathAlphabet{\mathbfit}{OT1}{cmr}{bx}{it}
  \SetMathAlphabet\mathbfit{bold}{OT1}{cmr}{bx}{it}
  \DeclareMathAlphabet{\mathbfss}{OT1}{cmss}{bx}{n}
  \SetMathAlphabet\mathbfss{bold}{OT1}{cmss}{bx}{n}
  \ifAMStwofonts
    \ifCUPmtlplainloaded \else
      \DeclareSymbolFont{UPM}{U}{eur}{m}{n}
      \SetSymbolFont{UPM}{bold}{U}{eur}{b}{n}
      \DeclareSymbolFont{AMSa}{U}{msa}{m}{n}
      \DeclareMathSymbol{\upi}{0}{UPM}{"19}
      \DeclareMathSymbol{\umu}{0}{UPM}{"16}
      \DeclareMathSymbol{\upartial}{0}{UPM}{"40}
      \DeclareMathSymbol{\leqslant}{3}{AMSa}{"36}
      \DeclareMathSymbol{\geqslant}{3}{AMSa}{"3E}
    \fi
  \fi
\fi 
 
\ifCUPmtlplainloaded \else
  \ifAMStwofonts \else 
    \def\upi{\pi}
    \def\umu{\mu}
    \def\upartial{\partial}
  \fi
\fi

\input psfig.sty
\input epsf.sty

\title[The Significance of the {\it Hipparcos} Orbit of $\rho$ CrB]
{On the Statistical Significance of the {\it Hipparcos} \\
Astrometric Orbit of $\rho$ Coronae Borealis}

\author[S. Zucker and T. Mazeh]
{Shay Zucker\thanks{E-mail: shay@wise.tau.ac.il}
 and Tzevi Mazeh\thanks{E-mail: mazeh@wise.tau.ac.il} \\
School of Physics and Astronomy, Raymond and Beverly Sackler \\
Faculty of Exact Sciences, Tel Aviv University, Tel Aviv, Israel}

\begin{document}

\maketitle

\begin{abstract}
Recently Gatewood, Han \& Black \shortcite{Gatewood2001} presented an analysis 
of the {\it Hipparcos} stellar positions and their own
ground-based measurements of \RCB, suggesting an astrometric
orbit of 1.5 \mas\ with an extremely small orbital inclination of
0.5\degr. This indicates that the
planet-candidate secondary might be a late M star.  
We used the {\it Hipparcos} data of \RCB\ {\it together} with the
individual radial velocities of Noyes et al. \shortcite{Noyes1997} to independently
study the stellar orbit and to assess its statistical significance. Our
analysis yielded the same astrometric orbit.  
However, a permutation
test we performed on the {\it Hipparcos} measurements indicated that the
statistical significance of the astrometric orbit is only
$2\sigma$. Therefore, we can expect about one out of 40 systems with a
similar data set 
to show a false astrometric orbit. 
\end{abstract}
\begin{keywords}
astrometry -- planetary systems -- 
stars: individual (\RCB) -- techniques: radial velocities
\end{keywords}

\section{Introduction}

One of the first stars around which a planet candidate was found \cite{Noyes1997}
was $\rho$ Coronae Borealis
(HR~5968 = HD~143761 = HIP~78459). The discovery was based on the detection
of a small periodic radial-velocity modulation, with an amplitude of 67
\ms. The corresponding minimum mass of the unseen companion was found to
be 1 Jupiter mass (=\MJ), its actual mass depending on the orbital
inclination.

Recently Gatewood, Han \& Black \shortcite{Gatewood2001} presented an analysis of
the {\it Hipparcos} intermediate data, together with new ground-based
astrometry they obtained with the Multichannel Astrometric Photometer
\cite{Gatewood1987}. Their analysis suggests an orbital semi-major 
axis of 1.5 \mas(=mas) with an extremely small orbital inclination --
0.5\degr, indicating that the secondary might be a late M star with a
mass of $0.14\pm 0.05$ \Mo. 
The extremely small orbital inclination makes this
discovery intriguing, because the probability of a binary to have such a
small angle, assuming an {\it isotropic} orientation in space, is
extremely small -- $4\times10^{-5}$.  Even among 50 systems, the
current number of planet candidates \cite{Schneider2000}, the probability to
find one system with such an extremely small inclination is only
$0.002$.  Therefore, the statistical significance of the \GHB\ analysis,
which might have implications beyond the study of \RCB, should be
estimated carefully.                                   

Recently, Pourbaix \shortcite{Pourbaix2001} studied the significance of the
derived {\it Hipparcos} orbits of many planet candidates, including \RCB. 
Pourbaix fitted an orbit to the {\it Hipparcos} data and checked the
improvement in the fit resulting from the additional parameters,
using an F-test. He used the F-distribution to assess the final
significance level of the derived orbit, concluding that the
significance is somewhat high -- 99 per cent.
This use of the F-distribution assumes Gaussianity
of the individual measurements. We avoid this assumption by using the
permutation test,
which belongs to the class of distribution-free tests
(e.g. Good 1994)
and thus is more robust against modeling problems of the
measurement process.

We present in this paper an independent analysis of
only the {\it Hipparcos} astrometric data (i.e. without the MAP data), with
an emphasis on the assessment of the
 significance of the detection. On one hand, we show that the {\it Hipparcos}
 data alone, together with the
 radial-velocity measurements of Noyes et al. \shortcite{Noyes1997}, yield, indeed, a
small astrometric orbit of 1.5 mas, implying a
 small inclination for \RCB. On the other hand, we show that the
 statistical significance of this finding is only about $2\sigma$. Therefore, we can
 expect about one system out of 40 to have shown a false astrometric
 orbit. Section 2 presents our analysis, while Section 3 discusses our
 finding.

\section{Analysis}

\subsection{Orbital Solution}

The present analysis used the 38 AFOE radial velocities of \RCB\ 
(Noyes et al. 1997; 1999),
together with the 42 astrometric measurements of
{\it Hipparcos} \cite{ESA1997}, which were analysed by the FAST and the NDAC
consortia \cite{vanLeeuwen1998}.  The spectroscopic and
astrometric solutions have in common the following elements: the period,
$P$; the time of periastron passage, $T_0$; the eccentricity, $e$; the
longitude of the periastron, $\omega$.  In addition, the spectroscopic
elements include the radial-velocity amplitude, $K$, and the
centre-of-mass radial velocity $\gamma$.  The astrometric orbital
elements include three additional elements -- the angular semi-major
axis of the photocentre, $a_0$; the inclination, $i$; the longitude of
the nodes, $\Omega$. In addition, the astrometric solution includes the
five regular astrometric parameters -- the parallax, the position (in
right ascension and declination) and the proper motion (in right
ascension and declination). All together we had a 14-parameter model to
fit to the spectroscopic and astrometric data. Note that we have
allowed small corrections to the values of the {\it Hipparcos} reference
solution of the five regular astrometric parameters.

These 14 elements are not all independent. From $K$, $P$ and $e$ we can
derive the projected semi-major axis of the primary orbit --
$a_{1,phys}\times \sin i$, in physical units. This element, together
with the inclination $i$ and the parallax, yields the angular semi-major
axis of the primary, $a_1$.  Assuming the secondary contribution to the
total light of the system is negligible, this is equal to the observed
$a_0$.

The results of our fit, which are consistent with the previous orbital
solutions 
(Noyes et al. 1997, 1999; Gatewood et al. 2001),
are given in Table \ref{Table1}. The small semi-major axis implies
a secondary mass of 
$
M_2=0.125 \pm 0.042 \ M_{\odot},
$
where we assumed 1 \Mo\ for the primary \cite{Noyes1997}. 

\begin{table}
\caption{Best-Fitting Orbital Solution for \RCB.}
\label{Table1}
\begin{tabular}{lcl}
\hline
$P$        &   $39.73       \pm 0.11$      & days  \\
$K$        &   $67.8        \pm 2.8$       & \ms\  \\
$e$        &   $0.023       \pm 0.047$     &       \\ 
$T_0$      &   $2,449,060   \pm 16 $       & HJD   \\
$\omega$   &   $7\degr      \pm 129\degr$  &       \\
$\gamma$   &   $47.8        \pm 2.2 $      & \ms\ \\
$a_1\sin i$&   $2.48        \pm 0.10$      & $10^{-4}$AU\\
\\
$i$        &   $0\fdg54   ^{+0\fdg24}_{-0\fdg13}$   & \\
$\Omega$   &   $118\degr     \pm 52\degr$  &       \\
$a_0$      &   $1.49        \pm 0.46$      & \mas\ \\
parallax   &   $56.70       \pm 0.63$      & \mas\ \\
\hline
\end{tabular}
\end{table}

\subsection{Significance}

The semi-major axis of the derived astrometric orbit is $1.49\pm
0.46$ mas. A very
small semi-major axis could have been falsely ``detected'' even without
any real astrometric motion, due to the scatter of the actual
measurements \cite{Halbwachs2000}.  To find out if this is the case
here we performed a ``permutation'' test (e.g., Good 1994) by
generating simulated data from the very same astrometric measurements of
\RCB. If there is some evidence of an orbit in the measurements, it should
be ruined by the permutation.
However, if the derived orbit
is spurious, some random permutations should be able to reproduce
a similar effect. In a sense, we let the data ``speak for themselves''
and do not have to assume any specific distribution for the measurements.

We used the IAD {\it Hipparcos} measurements \cite{ESA1997}
and permuted the actual timing of the
observations, modifying the partial derivatives with respect to the five
astrometric parameters \cite{ESA1997} accordingly.  We then analysed the
permuted astrometric data together with the actual radial velocities,
deriving a new false astrometric orbit.

For most of the original measurements there are two stellar positions,
one derived from the NDAC and the other from the FAST consortia. The two
positions are, obviously, not independent, but have an assigned 
non-vanishing correlation.  
In our permutation we kept the pairing of the corresponding NDAC and
FAST positions, while permuting the timings among the pairs.
This turned out to be a
nontrivial exercise, and caused us to falsely assign a higher
significance to the astrometric detection in an early phase of this study.

\begin{figure}
\epsfbox{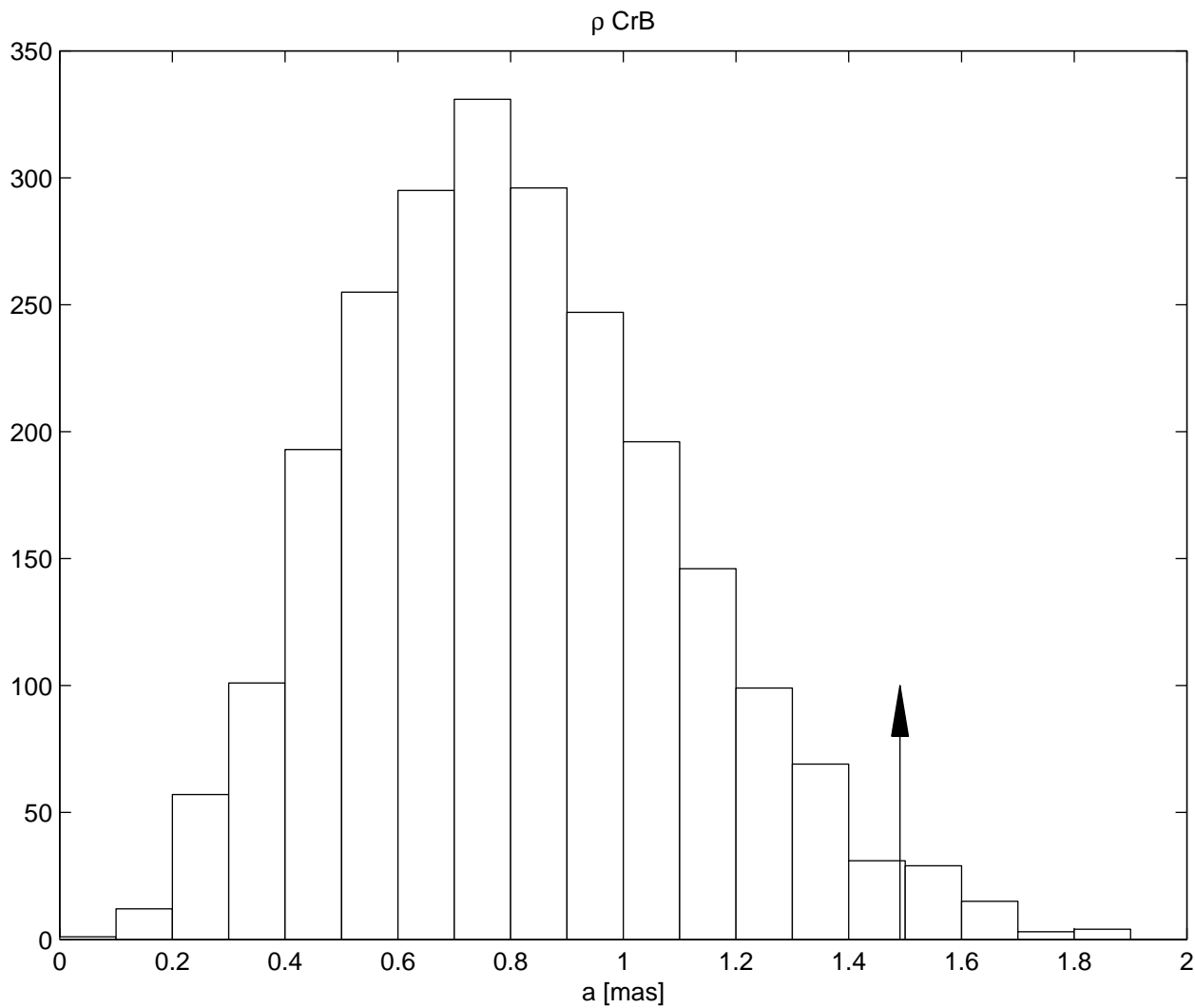}
\caption{Histogram of the size of the falsely ``detected'' semi-major
axes in the simulated permuted data. The size of the actually detected
axis is marked by an arrow.}
\label{Figure1}
\end{figure}

A histogram of the falsely detected semi-major axes, derived from the
``simulated'' permuted data, is presented in Figure \ref{Figure1}.  We see that most
of the artificially detected semi-major axes are at the range of
0.5--1.0 mas, with a small distribution tail beyond 1.5 mas.
In 2,000 random permutations we got 46 astrometric
solutions with semi-major axis larger than 1.5. This indicates that the
detection significance is at the level of 0.977, which is about
$2\sigma$. 

\section{Discussion}

The analysis presented here shows that on average we expect about one
out of 40 systems with a similar dataset to \RCB\ to show a false
astrometric 
orbit with $a=1.5$ mas. This 
implies that
the statistical significance of the claimed astrometric orbit is not
high enough 
at this point of the research.

As explained in the introduction, Pourbaix \shortcite{Pourbaix2001} 
derived that the significance of the astrometric
solution is at the 99 per cent level. Compared to his technique we use
a distribution-free analysis and thus we feel that our assertion, of a
97.7 per cent level, is more justified.

One way to find out the true nature of the secondary of \RCB\ is to try
and observe the secondary directly. If the secondary is indeed a stellar
object, infrared spectroscopic observations, especially when analysed
with TODCOR -- a two-dimensional correlation technique \cite{Zucker1994},
should detect some trace of the faint secondary (e.g., Mazeh et al. 2000a).
Together with M. Simon and L. Prato, we plan such
observations in the near future. Obviously, more precise astrometric
measurements, expected to come in the future from the FAME
\cite{Horner1999}, SIM \cite{NASA1999},
DIVA \cite{Roser1998} and GAIA \cite{Gilmore1998}
missions, will
enable us to derive the secondary mass much more accurately.

\begin{figure}
\epsfbox{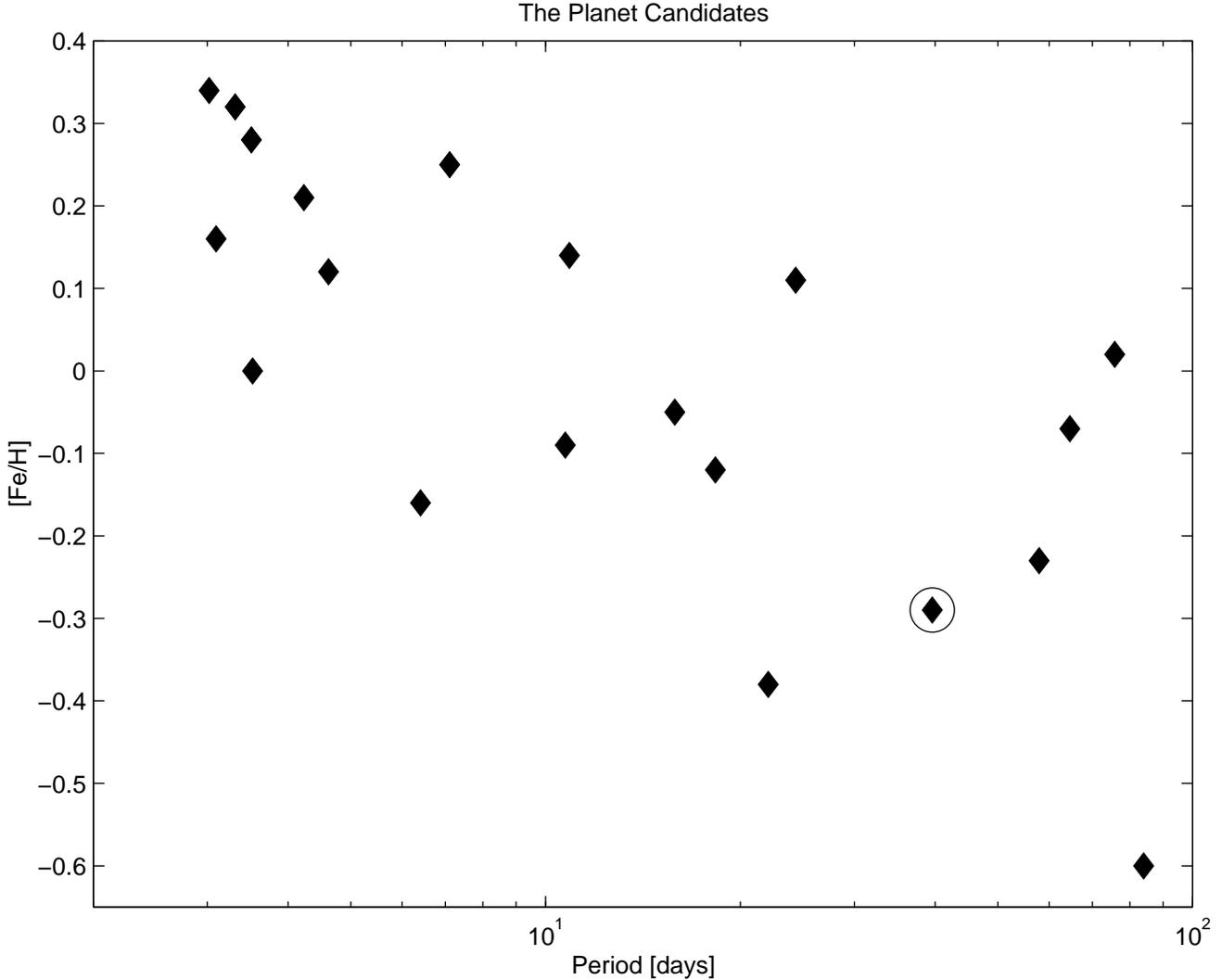}
\caption{The stellar metallicity as a function of the 
planet-candidate orbital period for planets with periods shorter than 100 days.
The point representing \RCB\ is circled.}
\label{Figure2}
\end{figure}

We turn now to discuss the relatively low metallicity of \RCB.  It
was noted by many workers (Gonzalez 1997; Marcy \& Butler 1998;
Queloz et al. 2000; Gonzalez 2001; Butler et al. 2000)
that most stars that
were found to harbor planet candidates exhibit metallicities higher than
the typical metallicity of the solar neighborhood. Queloz et al. \shortcite{Queloz2000}
and Butler et al. \shortcite{Butler2000} further pointed out that the host stars to the
``51 peg like'' planets are particularly metal rich. Mazeh \& Zucker \shortcite{Mazeh2001}
suggested a probable decrease of the stellar metallicity as a
function of the orbital period of the planet candidates, a dependence
which holds, according to their suggestion, up to about 100 days.  To
study their suggestion we plot in Figure \ref{Figure2} the metallicity as a
function of the planet orbital period, for the stars that appear in 
The Extra Solar Planets Encyclopedia web site as of September 2000 
\cite{Schneider2000}. We follow the Mazeh \& Zucker \shortcite{Mazeh2001}
suggestion and therefore include in the figure only planet
candidates with periods shorter than 100 days.

Whenever available we have used metallicities derived from spectral
analysis, mostly from the seminal work of Gonzalez \shortcite{Gonzalez2001}. 
Mazeh et al. \shortcite{Mazeh2000b}
derived the metallicity of HD~209458. Whenever such an
analysis was not available, we have used the photometric metallicity
derived by Gim\'{e}nez \shortcite{Gimenez2000}.  The metallicities of the stars not
considered by Gim\'{e}nez were derived by us following his prescription,
based on the photometry of Hauck \& Mermilliod \shortcite{Hauck1998} and the
calibrations calculated by Crawford \shortcite{Crawford1975} and 
Olsen \shortcite{Olsen1984}.  

In this figure \RCB\ has one of the lowest metallicities. 
The analysis presented here, which does not support the conjecture
about the stellar nature of the companion of \RCB, strengthens
the suggestion that the metallicity might depend on the period.
The figure draws attention to the other
two low-metallicity planet candidates. The extreme one is HD~114762
(Latham et al. 1989; Mazeh, Latham \& Stefanik 1996) with a metallicity
of about $-$0.60. 
The nature of this secondary is not clear because it 
has a minimum mass of about 10 \MJ, and
even a moderate inclination can turn it into a brown-dwarf or stellar
secondary. However, The other one is HD~6434 \cite{Udry2000}, with a minimum
mass of 0.5 \MJ. This small minimum mass renders the conjecture that this
is a real planet safe.
Therefore, even if the secondary of HD~114762 turns out to be stellar, the dependence 
of the metallicity on the period might still be real.

\section*{acknowledgments}
We are indebted to R. Noyes for many seminal
suggestions, which led to substantial improvement of this study.
M. Simon is thanked for his wise advise. 
This work was supported by
the US-Israel Binational Science Foundation through grant 97-00460 and
the ISRAELI SCIENCE FOUNDATION (grant no. 40/00).

\end{document}